\documentclass[
    ,final            % use final for the camera ready runs
 ,biblio
]{aipproc}
\usepackage{amsmath}

\def\vv{{\boldsymbol v}}

\def\vecr{{{\vec r}}}

\def\vxi{{\boldsymbol\xi}}

\layoutstyle{8x11double}

\newcommand{\aap}{    {\rm Astron. Astrophys.}\ }

\newcommand{\apj}{    {\rm Astrophys. J.}\ }
\newcommand{\apjl}{   {\rm Astrophys. J. Lett.}\ }

\newcommand{\mnras}{  {\rm Mon. Not. Roy. Astron. Soc.}\ }
\newcommand{\nat}{    {\rm Nature}\ }

\newcommand{\pasj}{   {\rm Pub. Astron. Soc. Japan}\ }

\newcommand{\solphys}{{\rm Solar Phys.}\ }

\begin{document}

\title{Solar Oscillations}

\classification{<Replace this text with PACS numbers; choose from
this list:
                \texttt{http://www.aip..org/pacs/index.html}>}
\keywords      {solar oscillations, helioseismology, solar interior}

\author{A. G. Kosovichev}{
  address={HEPL, Stanford University, Stanford, CA 94305}
}

\begin{abstract}
 In recent years solar oscillations have been studied in great
detail, both observationally and theoretically; so, perhaps, the Sun
currently is the best understood pulsating star. The observational
studies include long, almost uninterrupted series of oscillation
data from the SOHO spacecraft and ground-based networks, GONG and
BiSON, and more recently, extremely high-resolution observations
from the Hinode mission. These observational data cover the whole
oscillation spectrum, and have been extensively used for
helioseismology studies, providing frequencies and travel times for
diagnostics of the internal stratification, differential rotation,
zonal and meridional flows, subsurface convection and sunspots.
Together with realistic numerical simulations they lead to better
understanding of the excitation mechanism and interactions of the
oscillations with turbulence and magnetic fields. However, many
problems remain unsolved. In particular, the precision of the
helioseismology measurements is still insufficient for detecting the
dynamo zone and deep routes of sunspots. Our knowledge of the
oscillation physics in strong magnetic field regions is inadequate
for interpretation of MHD waves in sunspots and for sunspot
seismology. A new significant progress in studying the solar
oscillations is expected from the Solar Dynamics Observatory
scheduled for launch in 2010.
\end{abstract}

\maketitle

%%%%%%%%%%%%%%%%%%%%%%%%%%%%%%%%%%%%%%%%%%%%
%% MAINMATTER
%%%%%%%%%%%%%%%%%%%%%%%%%%%%%%%%%%%%%%%%%%%%

\section{Introduction}
Solar oscillations have been studied extensively since their
discovery by \citet{Leighton1962}. Currently, it is well-understood
that the 5-min oscillations that dominate the dynamics of the solar
atmosphere represent radial and non-radial acoustic and surface
gravity modes stochastically excited by turbulent convection and
trapped below the surface. In addition, impulsive localized
excitation by solar flares has been observed in several cases. The
solar oscillation modes are observed in a wide range of angular
degree, from 0 to 3000, and used to infer the internal structure and
dynamics of the Sun by methods of helioseismology. These methods are
generally divided into two classes: global and local
helioseismology.
\begin{figure}[h]
  \includegraphics[height=.4\textheight]{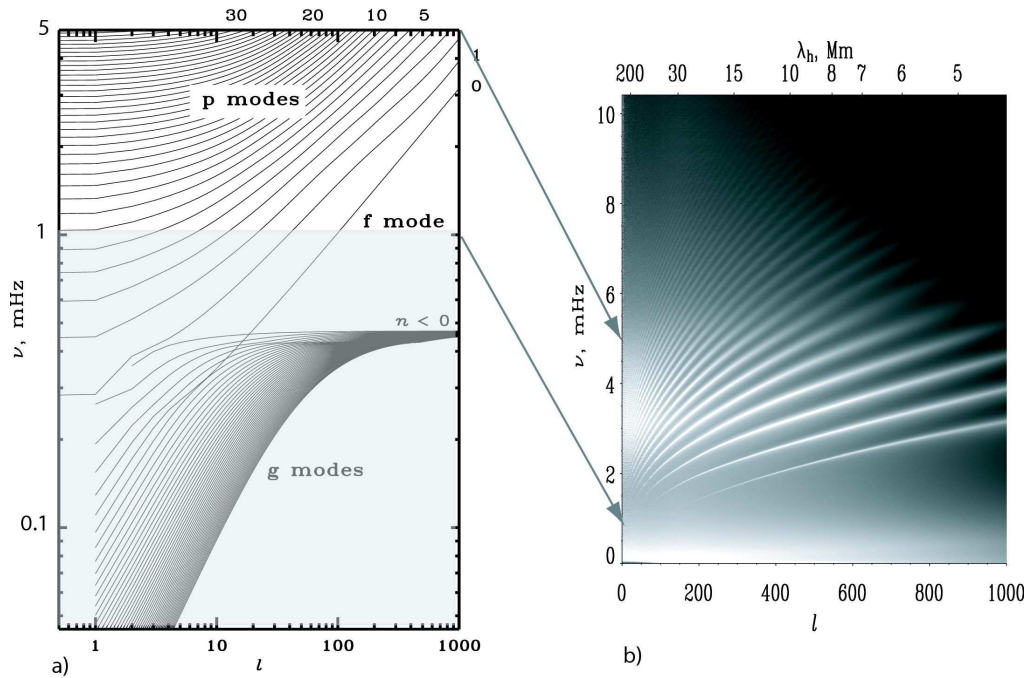}
  \caption{a) Theoretical spectrum of solar oscillations \citep{JCD2002}.
b) The power spectrum of solar oscillations observed from SOHO/MDI.}
\end{figure}

Global helioseismology uses measurements of normal mode frequencies
and is based on the classical theory of stellar oscillations
\citep{Ledoux1958}. This theory provides relationships between
normal modes frequencies and interior properties such as the radial
and latitudinal distributions of the sound speed, density and
angular velocity. Local helioseismology measures local 3D
perturbations of the sound speed and flow fields associated with
large-scale solar convection and magnetic structures. These methods
are based on measurements of wave dispersion properties (e.g.
frequency shifts in local areas) and wave travel times. Theoretical
description of these techniques is more complicated than for the
global methods, and is still being developed. The main complexity
comes from the need to take into account coupling among the normal
modes caused by 3D perturbation. Global helioseismology analyses
generally ignore this coupling and consider only frequency shifts
and splitting caused by spherical and axisymmetrical (and also
North-south symmetrical) perturbations. While the axisymmetrical
approximation only crudely describes the structure and dynamics of
the Sun the global helioseismology methods are currently the main
tool for probing the deep interior and variations in the convection
zone associated with the solar activity cycle. Three-dimensional
local helioseismology methods are still limited to diagnostics of
relatively shallow subsurface layers. However, some initial attempts
have been made to probe the structure of the tachocline (a
transition layer between the radiative core and the convective
envelope) by a time-distance helioseismology method
\citep{Zhao2009}.

While the physics of solar oscillations and their mechanism are
generally understood many interesting and important details are
still unknown. Among these are the precise nature of turbulent
perturbations that drive solar oscillations, mechanisms of asymmetry
of the oscillation line profiles in the observed power spectrum,
phase and amplitude relations between velocity and intensity
oscillations, the role of magnetic fields in the mode excitation and
properties, as well as non-adiabatic effects caused by interaction
of waves with turbulence and radiation. Also, the mechanism of
flare-excited oscillation ("sunquakes") is not well-understood.

Knowing the physics of solar oscillations is very important for
improving the accuracy of helioseismic measurements. For instance,
initially, frequencies of normal modes were measured by fitting a
symmetrical Lorentzian profile to the  peaks in the oscillation
power spectra. This profile comes from a simplified model that
treats the oscillation modes as a damping harmonic oscillator (e.g.
\cite{Chaplin2005}). However, this results in systematic frequency
shifts because the lines are asymmetrical. Thus, more recent
measurements are carried out using asymmetrical profiles
\citep{Nigam1998b}. Also, accurate models of the excitation source
are required for calculating sensitivity functions for acoustic
travel times using a Born approximation in time-distance
helioseismology \citep{Birch2000,Birch2004}. For this type of
measurements it is particularly important to take into account
variations of the strength and spectral distribution in magnetic
regions. In an extreme case of sunspot umbra the excitation of
acoustic waves is suppressed because the strong magnetic field
inhibits convection. The spatial and spectral variations of the
oscillation power may cause systematic shifts in travel-time
measurements when a phase-speed filtering procedure to improve the
signal-to-noise ratio is applied
\citep{Rajaguru2006,Parchevsky2008}. Thus, the current research in
the field of solar oscillations is focused on improving the
understanding of the excitation mechanism, effects of magnetic
fields on wave excitation, propagation and damping, and also on
developing more accurate methods of helioseismology, particularly,
the local techniques. This article presents a brief overview of some
aspects of the solar oscillation physics and helioseismology
methods, but no means this is a comprehensive literature review.

\section{Physics of Solar Oscillations}

\subsection{Oscillation Power Spectrum}

The theoretical spectrum of solar oscillation modes shown in Fig.~1a
covers a wide range of frequencies and angular degrees. It includes
oscillations of three types: acoustic (p) modes, surface gravity (f)
modes and internal gravity (g) modes. In this spectrum each curve
corresponds to a specific overtone of non-radial modes, which can be
described by the number of nodes along the radius (or by the radial
order, $n$). The angular degree, $l$, of corresponding spherical
harmonics describes the horizontal wave number (or inverse
horizontal wavelength). The p-modes cover the frequency range from
0.3 to 5 mHz (or from 3 to 55 min in oscillation periods). The low
frequency limit corresponds to the first radial harmonic, and the
upper limit is set by the acoustic cut-off frequency of the solar
atmosphere. The g-modes have an upper limit corresponding to the
maximum Br\"unt-V\"ais\"al\"a frequency ($\sim 0.45$ mHz) in the
radiative zone and occupy the low-frequency part of the spectrum.
The intermediate frequency range of 0.3-0.4 mHz at low angular
degrees is a region of mixed modes. These modes behave like g-mode
in the deep interior and like p-modes in the outer region. The
apparent crossings in this diagram are not the actual crossings: the
mode branches become close in frequencies but do not cross. A
similar phenomenon is known in quantum mechanics as avoided
crossing.
\begin{figure}
  \includegraphics[height=.4\textheight]{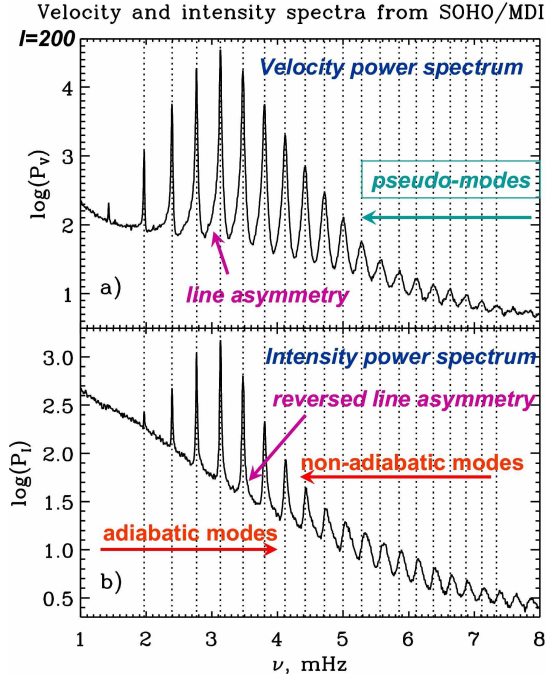}
  \caption{Power spectra of $l=200$ mode obtained from SOHO/MDI
observations of a) Doppler velocity, b) continuum intensity.
\citep{Nigam1998a} }
\end{figure}

So far, only the upper part of the solar oscillation spectrum is
observed. The lowest frequencies of detected p- and f-modes are of
about 1 mHz. Below this frequency the mode amplitudes decrease below
the noise level, and become unobservable. There have been several
attempts to identify low-frequency p-modes or even g-modes in the
noisy spectrum, but sofar these results are not convincing.

The observed power spectrum is shown in Fig.~1b. The lowest ridge is
the f-mode, and the other ridges are p-modes of the radial order,
$n$, starting from $n=1$. The ridges of the oscillation modes
disappear in the convective noise at frequencies below 1 mHz. At low
angular degrees only high-$n$ modes are observed. However, the
$n$-values of these modes can be easily determined by tracing the
the high-$n$ ridges of the high-degree modes into the low-degree
region. This provides unambiguous identification of the low-degree
solar modes. Obviously, the mode identification is much more
difficult for spatially unresolved oscillations of other stars.
\begin{figure}
  \includegraphics[height=.3\textheight]{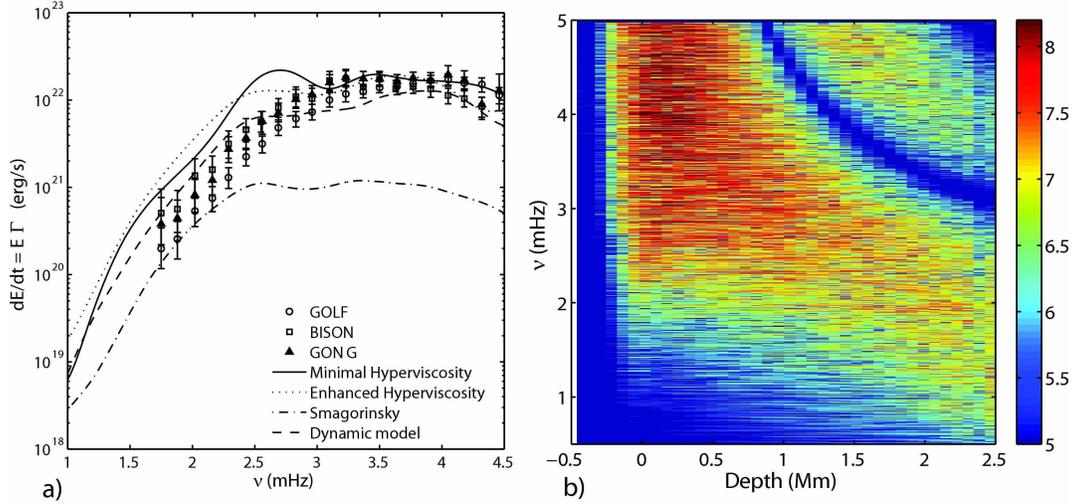}
  \caption{a) Comparison of observed and calculated rate of stochastic
energy input to modes for the entire solar surface ($erg.s^{-1}$).
Observed distributions: $circles$ SoHo-GOLF, $squares$ BISON, and
$triangles$ GONG for $l=1$ \citep{Baudin2005}. b) Logarithm of the
work integrand in units of $erg.cm^{-2}.s^{-1}$), as a function of
depth and frequency for numerical simulations with the dynamic
turbulence model. \citep{Jacoutot2008a}}
\end{figure}

\subsection{Excitation Mechanism}

Solar oscillations are driven by turbulent convection in a shallow
subsurface superadiabatic layer where convective velocities are the
highest. However, details of the stochastic excitation mechanism are
not fully established. Solar convection in the superadiabatic layer
forms small-scale granulation cells. Analysis of observations and
numerical simulations has shown that sources of solar oscillations
are associated with strong downdrafts in dark intergranular lanes
\citep{Rimmele1995}. These downdrafts are driven by radiative
cooling and may reach near-sonic velocity of several km/s. This
process has features of convective collapse \citep{Skartlien2000}.

Calculations of the work integral for acoustic modes using the
realistic numerical simulations of Stein and Nordlund
\citep{Stein2001} have shown that the principal contribution to the
mode excitation is provided by turbulent Reynolds stresses and that
a smaller contribution comes from non-adiabatic pressure
fluctuations.

As we have pointed out observations show that the modal lines in the
oscillation power spectrum are not Lorentzian but display a strong
asymmetry \cite{Duvall1993,Toutain1998}. Curiously, the asymmetry
has the opposite sense in the power spectra calculated from Doppler
velocity and intensity oscillations. The asymmetry itself can be
easily explained by interference of waves emanated by a localized
source \citep{Gabriel1992}, but the asymmetry reversal is surprising
and indicates on complicated radiative dynamics of the excitation
process, and is still not fully understood. However, it is clear
that the line shape of the oscillation modes and the phase-amplitude
relations of the velocity and intensity oscillations carry
substantial information about the excitation mechanism and, thus,
require careful data analysis and modeling.

\subsection{Line Asymmetry and Pseudo-modes}

Figure 2 shows the oscillation power spectrum of $l=200$ mode,
obtained from the SOHO/MDI Doppler velocity and intensity data. The
line asymmetry is apparent, particularly, at low frequencies. In the
velocity spectrum, there is more power in the low-frequency wings
than in the high-frequency wings of the spectral lines. In the
intensity spectrum the distribution of power is reversed. The data
also show that the asymmetry varies with frequency. It is the
strongest for the f-mode and low-frequency p-mode peaks. At higher
frequencies the peaks become more symmetrical, and extend well above
the acoustic cut-off frequency, which is about 5 mHz.

Acoustic waves with frequencies below the cut-off frequency are
completely reflected by the surface layers because of the steep
density gradient. These waves are trapped in the interior, and their
frequencies are determined by the resonant conditions, which depend
on the solar structure. But the waves with frequencies above the
cut-off frequency escape into the solar atmosphere. Above this
frequency the power spectrum peaks correspond to so-called
"pseudo-modes". These are caused by constructive interference of
acoustic waves excited by the sources located in the granulation
layer traveling upward and the waves traveling downward, reflected
in the deep interior and arriving back to the surface. Frequencies
of these modes are no longer determined by the resonant conditions
of the solar structure. They depend on the location and properties
of the excitation source ("source resonance"). The pseudo-mode peaks
in the velocity and intensity power spectra are shifted relative to
each other by almost a half-width. They are also slightly shifted
relative to the normal mode peaks despite they look like a
continuation of the normal-mode ridges in Figs~1b and 4a. This
happens because the excitation sources are located in a shallow
subsurface layer, which is very close to the reflection layers of
the normal modes. Changes in the frequency distributions below and
above the acoustic cut-off frequency can be easily noticed by
plotting the frequency differences along the modal ridges.
\begin{figure}
  \includegraphics[height=.5\textheight]{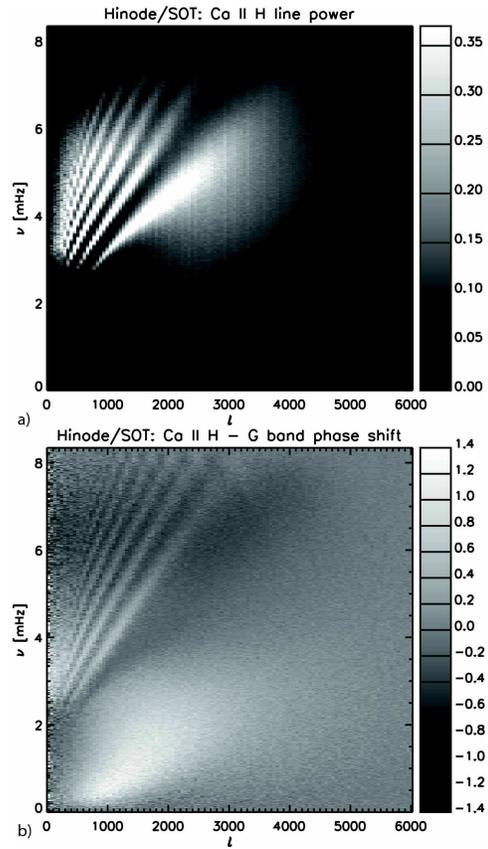}
  \caption{a) The oscillation power spectrum from Hinode CaII H line
  observations. b) The phase shift between CaII H and G-band
(units are in radians). \citep{Mitra-Kraev2008}}
\end{figure}
\begin{figure}
  \includegraphics[height=.2\textheight]{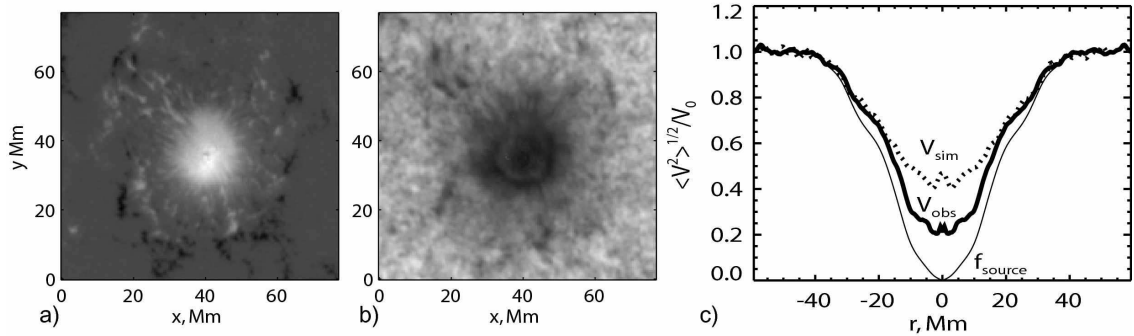}
  \caption{a) Line-of-sight magnetic field map of a sunspot (AR8243); b) oscillation amplitude map; c)
profiles of rms oscillation velocities at frequency 3.65 mHz for
observations (thick solid curves) and simulations (dashed curves);
the thin solid curve shows the distribution of the simulated source
strength.\citep{Parchevsky2007b}}
\end{figure}

The asymmetrical profiles of normal-mode peaks are also caused by
the localized excitation sources. The interference signal between
acoustic waves traveling from the source upwards and  the waves
traveling from the source downward and coming back to the surface
after the internal reflection depends on the wave frequency.
Depending on the source (multipole) type the interference signal can
be stronger at frequencies lower or higher the resonant normal
frequencies, thus resulting in asymmetry in the power distribution
around the resonant peak. Calculations of Nigam et al
\citep{Nigam1998a} showed that the asymmetry observed in the
velocity spectra and the distribution of the pseudo-mode peaks can
be explained by a composite source consisting of a monopole term
(mass term) and a dipole term (force due to Reynolds stress) located
in the zone of superadiabatic convection at a depth of $75\pm 50$ km
below the photosphere. In this model the reversed asymmetry in the
intensity power spectra is explained by effects of a correlated
noise added to the oscillation signal through fluctuations of solar
radiation during the excitation process. Indeed, if the excitation
mechanism is associated with the high-speed turbulent downdrafts in
dark lanes of granulation the local darkenings contribute to the
intensity fluctuations caused by excited waves. The model also
explains the shifts of pseudo-mode frequency peaks and their higher
amplitude in the intensity spectra.

While this scenario looks plausible and qualitatively explains the
main properties of the power spectra details of the physical
processes are still uncertain. In particular, it is unclear whether
the correlated noise affects only the intensity signal or both the
intensity and velocity. It has been suggested that the velocity
signal may have a correlated contribution due to convective
overshoot \citep{Roxburgh1997}. Attempts to estimate the correlated
noise components from the observed spectra have not provided
conclusive results \citep{Severino2001,Wachter2005}. Realistic
numerical simulations \citep{Georgobiani2003} have reproduced the
observed asymmetries and provided an indication that radiation
transfer plays a critical role in the asymmetry reversal.

Recent high-resolution observations of solar oscillation
simultaneously in two intensity filters, in molecular G-band and
CaII H line, from the Hinode space mission
\citep{Kosugi2007,Tsuneta2008} revealed significant  shifts in
frequencies of pseudo-modes observed in the CaII H and G-band
intensity oscillations \citep{Mitra-Kraev2008}. The phase of the
cross-spectrum of these oscillations shows peaks associated with the
p-mode lines but no phase shift for the f-mode (Fig.~4b). The p-mode
properties can be qualitatively reproduced in a simple model with a
correlated background if the correlated noise level in the Ca II H
data is higher than in the G-band data \citep{Mitra-Kraev2008}.
Perhaps, the same effect can explain also the frequency shift of
pseudo-modes. The CaII H line is formed in the lower chromosphere
while the G-band signal comes from the photosphere. But how this may
lead to different levels of the correlated noise is unclear.
\begin{figure}
  \includegraphics[height=.25\textheight]{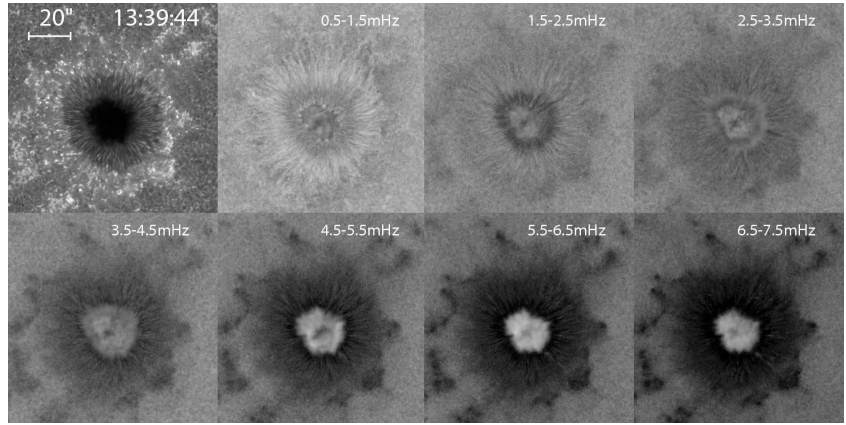}
  \caption{CaII H intensity image from Hinode observations
(top-left) and the corresponding power maps from CaII H intensity
data in five frequency intervals of active region NOAA 10935. The
field of view is 100 arcsec square in all the panels. The power is
displayed in logarithmic greyscaling. \citep{Nagashima2007}.}
\end{figure}

The Hinode results suggest that multi-wavelength observations of
solar oscillations, in combination with the traditional
intensity-velocity observations, may help to determine the level of
the correlated background noise and to determine the type of wave
excitation sources on the Sun.

In addition, Hinode provided observations of non-radial acoustic and
surface gravity modes of very high angular degree. These
observations show that the oscillation ridges are extended up to $l
\simeq 4000$ (Fig.~4a). In the high-degree range, $l \geq 2500$
frequencies of all oscillations exceed the acoustic cut-off
frequency. The line width of these oscillations dramatically
increases, probably, due to strong scattering on turbulence
\citep{Duvall1998}. Nevertheless, the ridge structure extending up
to 8 mHz (Nyquist frequency of these observations) is quite clear.
Although the ridge slope clearly changes at the transition from the
normal modes to the pseudo-modes.

\subsection{Magnetic Effects}

In general, the main factors causing variations in oscillation
properties in magnetic regions, can be divided in two types: direct
and indirect. The direct effects are due to additional magnetic
restoring forces that can change the wave speed and may transform
acoustic waves into different types of MHD waves. The indirect
effects are caused by changes in convective and thermodynamic
properties in magnetic regions. These include depth-dependent
variations of temperature and density, large-scale flows, and
changes in wave source distribution and strength. Both direct and
indirect effects may be present in observed properties such as
oscillation frequencies and travel times, and often cannot be easily
disentangled by data analyses causing confusions and
misinterpretations. Also, one should keep in mind that simple models
of MHD waves derived for various uniform magnetic configurations and
without stratification may not provide correct explanations to solar
phenomena. In this situation, numerical simulations play an
important role in investigations of magnetic effects.

Observed changes of oscillation amplitude and frequencies in
magnetic regions are commonly explained as a result of wave
scattering and conversion into various MHD modes. However, recent
numerical simulations helped to understand that magnetic fields not
only affect the wave dispersion properties but also the excitation
mechanism. In fact, changes in excitation properties of turbulent
convection in magnetic regions may play a dominant role in observed
phenomena.
\begin{figure}
  \includegraphics[height=.25\textheight]{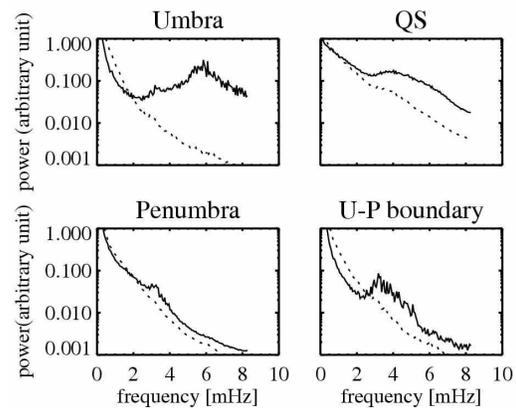}
  \caption{Power spectra averaged in the umbra (upper left),
in the quiet region (upper right), in the penumbra (lower left), and
around the boundary between the umbra and the penumbra (lower
right). The CaII H (solid) and the G-band (dotted) intensity power
spectra are shown. The ordinate is in an arbitrary unit in
logarithmic scale.\citep{Nagashima2007}}
\end{figure}

\subsubsection{Sunspot oscillations}

For instance, it is well-known that the amplitude of 5-min
oscillations is substantially reduced in sunspots. Observations show
that more waves are coming into the sunspot than going out of the
sunspot area (e.g. \citep{Braun1987}). This is often attributed to
absorption of acoustic waves in magnetic field due to conversion
into slow MHD modes traveling along the field lines (e.g.
\citep{Cally2009}). However, since convective motions are inhibited
by strong magnetic field of sunspots the excitation mechanism is
also suppressed. Three-dimensional numerical simulations of this
effect have shown that the reduction of acoustic emissivity can
explain at least 50\% of the observed power deficit in sunspots
(Fig.~5) \citep{Parchevsky2008}.

Another significant contribution comes from the amplitude changes
caused by variations in the background conditions. Inhomogeneities
in the sound speed may increase or decrease the amplitude of
acoustic wave traveling through these inhomogeneities. Numerical
simulations of MHD waves using magnetostatic sunspot models show
that the amplitude of acoustic waves traveling through sunspot
decreases when the wave is inside sunspot and then increases when
the wave comes out of sunspot \citep{Parchevsky2009}. Simulations
with multiple random sources show that these changes in the wave
amplitude together with the suppression of acoustic sources can
completely explain the observed deficit of the power of 5-min
oscillations. Thus, the role of the MHD mode conversion may be
insignificant for explaining the power deficit of 5-min photospheric
oscillations in sunspots. However, the mode conversion is expected
to be significant higher in the solar atmosphere where magnetic
forces become dominant.

We should note that while the 5-min oscillations in sunspots come
mostly from outside sources there are also 3-min oscillations, which
are probably intrinsic oscillations of sunspots. The origin of these
oscillations is not yet understood. They are probably excited by a
different mechanism operating in strong magnetic field.

Hinode observations added new puzzles to sunspot oscillations.
Figure~6 shows a sample Ca\,II\,H intensity and the relative
intensity power maps averaged over 1 mHz intervals in the range from
1 mHz to 7 mHz with logarithmic greyscaling \citep{Nagashima2007}.
 In the Ca\,II\,H power maps, in all the frequency ranges,
there is a small area ($\sim$ 6 arcsec in diameter) near the center
of the umbra where the power was suppressed.  This type of `node'
has not been reported before. Possibly, stable high-resolution
observation made by Hinode/SOT was required to find such a tiny
node, although analysis of other sunspots indicates that probably
only a particular type of sunspots, e.g., round ones with
axisymmetric geometry, exhibit such node-like structure.
 Above 4 mHz in the Ca\,II\,H power maps, power in the
umbra is remarkably high. In the power maps averaged over narrower
frequency range (0.05 mHz wide, not shown), the region with high
power in the umbra seems to be more patchy. This may correspond to
elements of umbral flashes, probably caused by overshooting
convective elements \citep{Schussler2006}. The Ca\,II\,H power maps
show a bright ring in the penumbra at lower frequencies. It probably
corresponds to the running penumbral waves.

Figure~7 shows the power spectra of the G-band and the Ca\,II\,H
intensity oscillations averaged in the quiet region, and sunspot
penumbra, umbra and a transition area between the umbra and penumbra
In all these regions, the G-band intensity power decreases almost
monotonically with the frequency, except the broad peak around 4 mHz
in the quiet region. This peak corresponds to the global five-minute
oscillation. The Ca\,II\,H intensity power in the quiet region shows
the similar trend, while the power spectrum in the penumbra exhibits
monotonic decrease, as is expected from the power maps (Fig.~6),
except the narrow peak at 3 mHz. The Ca\,II\,H intensity power
spectrum in the umbra has two peaks: one around 3 mHz and the other
around 5.5 mHz. In the previous works (e.g, a review by
\cite{Lites1992}, the dominant period of oscillation in the
chromosphere was above 5.5 mHz, and, in contrast to our results, no
significant power peaks were found in the 3 mHz range. The broad
peak in the umbra-penumbra transition region, between 2 mHz and 5
mHz, is caused by running penumbral waves.

\begin{figure}
  \includegraphics[height=.3\textheight]{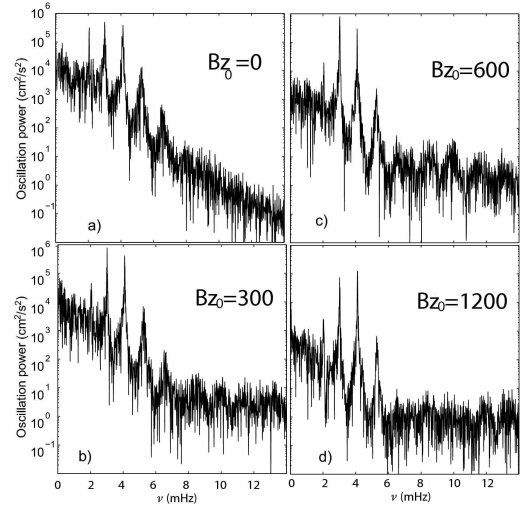}
  \caption{Power spectra of the horizontally averaged vertical velocity at the visible surface for
different initial vertical magnetic fields. The peaks on the top of
the smooth background spectrum of turbulent convection represent
oscillation modes: the sharp asymmetric peaks below 6 mHz are
resonant normal modes, while the broader peaks above 6 mHz, which
become stronger in magnetic regions, correspond to
pseudo-modes.\citep{Jacoutot2008b}}
\end{figure}
\begin{figure}
  \includegraphics[height=.4\textheight]{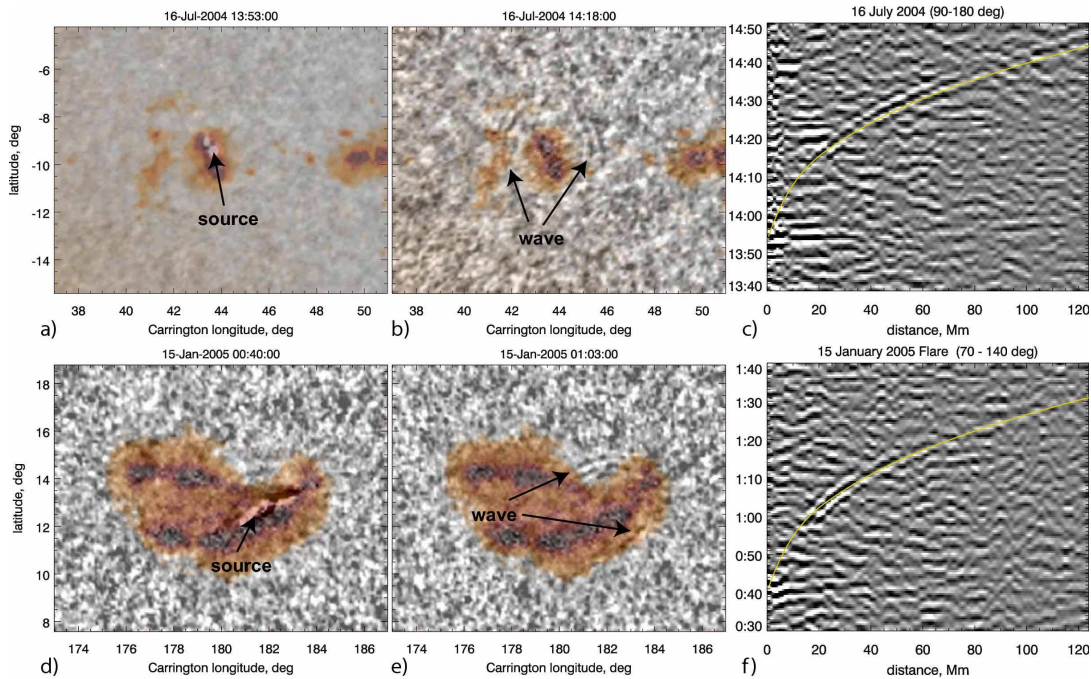}
  \caption{Observations of the seismic response of the Sun (``sunquakes'')
to two solar flares: a-c)  X3 of 16 July, 2004, and d-f) X1 flare of 15 January, 2005.
The left panels show a superposition of MDI white-light images of
the active regions and locations of the sources of the seismic waves
determined from MDI Dopplergrams,  the middle column shows the
seismic waves, and the right panels show the time-distance diagrams
of these events. The thin yellow curves in the right panels
represent a theoretical time-distance relation for helioseismic
waves for a standard solar model.\citep{Kosovichev2006b}}
\end{figure}

\subsubsection{Acoustic Halos}
In moderate field regions, such as plages around sunspot regions
observations reveal enhanced emission at high frequencies, 5-7 mHz,
(with period $\sim 3$ min) \citep{Braun1992}. Sometimes this
emission is called "acoustic halo". Radiative MHD simulations of
solar convection \citep{Jacoutot2008b} in the presence of vertical
magnetic field have shown that the magnetic field significantly
changes the structure and dynamics of granulations, and thus the
conditions of wave excitation. In magnetic field the granules become
smaller, and the turbulence spectrum is shifted towards higher
frequencies. This is illustrated in Figure~8, which shows the
frequency spectrum of the horizontally averaged vertical velocity.
Without magnetic field the turbulence spectrum declines sharply at
the frequencies above 5 mHz, but in the presence of magnetic field
it develops a plato. In the plato region characteristic peaks
(corresponding to the "pseudo-modes") appear in the spectrum for
moderate magnetic field strength of about 300-600 G. These peaks may
explain the effect of "acoustic halo". Of course, more detailed
theoretical and observational studies are required to confirm this
mechanism. In particular, multi-wavelength observations of solar
oscillations at several different heights would be important.
Investigations of the excitation mechanism in magnetic regions is
also important for interpretation of the variations of the frequency
spectrum of low-degree modes on the Sun, and for asteroseismic
diagnostics of stellar activity.

\subsection{Sunquakes}

\subsubsection{Excitation of Acoustic Waves by Flare Impact}

 ``Sunquakes", the helioseismic response to solar flares, are
caused by strong localized hydrodynamic impacts in the photosphere
during the flare impulsive phase. The helioseismic waves are
observed directly as expanding circular-shaped ripples in SOHO/MDI
Dopplergrams, which can be detected in Dopplergram movies and as a
characteristic ridge in time-distance diagrams,
\citep{Kosovichev1998, Kosovichev2006a}, or indirectly by
calculating integrated acoustic emission \citep{Donea1999,
Donea2005}. Solar flares are sources of high-temperature plasma and
strong hydrodynamic motions in the solar atmosphere. Perhaps, in all
flares such perturbations generate acoustic waves traveling through
the interior. However, only in some flares the impact is
sufficiently localized and strong to produce the seismic waves with
the amplitude above the convection noise level. It has been
established in the initial July 9, 1996, flare observations
\citep{Kosovichev1998} that the hydrodynamic impact follows the hard
X-ray flux impulse, and hence, the impact of high-energy electrons.

A characteristic feature of the seismic response in this flare and
several others \citep{Kosovichev2006a, Kosovichev2006b} is
anisotropy of the wave front: the observed wave amplitude is much
stronger in one direction than in the others. In particular, the
seismic waves excited during the October 28, 2003, flare had the
greatest amplitude in the direction of the expanding flare ribbons.
The wave anisotropy was attributed to the moving source of the
hydrodynamic impact, which is located in the flare ribbons
\citep{Kosovichev2006a, Kosovichev2006c}. The motion of flare
ribbons is often interpreted as a result of the magnetic
reconnection processes in the corona. When the reconnection region
moves up it involves higher magnetic loops, the footpoints of which
are further apart. Of course, there might be other reasons for the
anisotropy of the wave front, such as inhomogeneities in
temperature, magnetic field, and plasma flows. However, the source
motion seems to be quite important.

Therefore, we conclude that the seismic wave was generated not by a
single impulse  but by a series of impulses, which produce the
hydrodynamic source moving on the solar surface with a supersonic
speed. The seismic effect of the moving source can be easily
calculated by convolving the wave Green's function with a moving
source function. The results of these calculations a strong
anisotropic wavefront, qualitatively similar to the observations
\cite{Kosovichev2007a}. Curiously, this effect is quite similar to
the anisotropy of seismic waves on Earth, when the earthquake
rupture moves along the fault. Thus, taking into account the effects
of multiple impulses of accelerated electrons and moving source is
very important for sunquake theories. The impulsive sunquake
oscillations provide unique information about interaction of
acoustic waves with sunspots. Thus, these effects must be studied in
more detail.
\subsubsection{Excitation of Global Oscillations by Sunquakes and
Starquakes}

An interesting question is if solar and stellar flares can excite
global (low-degree) modes of the amplitude detectable by the current
telescopes. Baudin and Finidori \cite{Baudin2004} using the
 SOHO/GOLF Doppler-shift oscillation data found during the flare of July 9, 1996,
when the first "sunquake" was observed in the MDI data the power of
solar low-degree modes increased in the frequency range of $3.3-3.7$
mHz. However, a similar analysis of a stronger flare of October 28,
 2003, did not find a significant power increase that would indicate
on flare-excited global oscillations (Baudin, private
communication). Contrary, a surprisingly strong correlation between
increases of the whole-Sun acoustic power in the range of 5-7 mHz
observed from the SOHO/VIRGO total irradiance data and the solar
soft X-ray flux was found by Karoff and Kjeldsen \cite{Karoff2008}.
They interpreted this correlation as an evidence that solar flares
drive global oscillations of the Sun.
\begin{figure}
  \includegraphics[height=.5\textheight]{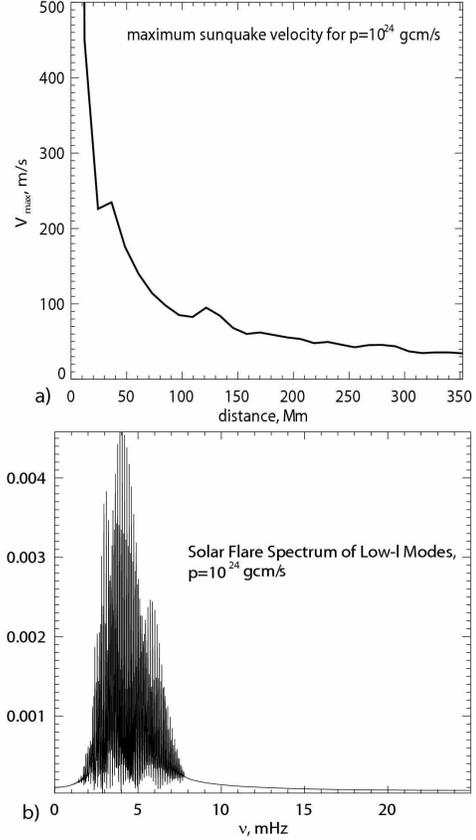}
  \caption{a) Theoretical spectrum of solar oscillations excited by an impulsive source
of the total momentum of $10^{24}$ g$\,$cm/s. b) The velocity
amplitude of low-degree modes $l=0-2$ excited by this source.}
\end{figure}

To resolve this controversy we estimate an upper amplitude of
low-degree solar modes excited by an impulsive impact at the solar
surface. The limits on the total momentum and energy of this source
can obtained from the spatially resolved observations of sunquakes.

Indeed, a solution of the non-radial stellar pulsations equation
written in symbolic form for a displacement vector $\vxi$:
  \[\frac{\partial^2\vxi}{\partial  t^2}+{\cal L}\vxi=0 \]
for initial conditions: $\vv(r,\theta,\phi,0)=\vv_0(r,\theta,\phi) $
can be obtained in terms of normal mode eigenfunctions, $vxi_{nlm}$:
 \[\vv(r,\theta,\phi,t)=\sum_{nlm}\frac{\langle\vxi_{nlm}^*\cdot\vv_0\rangle}
{\langle\vxi_{nlm}^*\cdot\vxi_{nlm}\rangle}
\vxi_{nlm}\cos(\omega_{nlm}t)e^{-\gamma_{nlm}t},\] where the angular
brackets mean the integration over the solar mass. Assuming that the
impact is localized in a very small volume at the surface and
calculating the integrals we obtain for the radial component of the
oscillation velocity at the surface:
\[ v_r(R,\theta,\phi,t)=\sum_{nlm}\frac{P_0}{M_\odot I_{nl}}
(2\ell+1)P_\ell(\cos\theta)\cos(\omega_{nl}t)e^{-\gamma_{nl}t}, \]
where $P_0$ is the total momentum of the impact, $M_\odot$ is the
solar mass, $I_{nl}$ is the mode inertia, $P_\ell$ is the Legendre
polynomial, $\omega_{nl}$ are the mode eigenfrequencies, and
$\gamma_{nl}$ are their damping times. The denominator, $M_\odot
I_{nl}$, is often "mode mass". Thus, the mode amplitude excited by
the impulsive point source is equal to the total momentum divided by
the mode mass and multiplied by a geometrical factor
$(2\ell+1)P_\ell$.

Using this solution we calculate the maximum amplitude of the wave
as a function of distance $R\theta$ from the impact (Fig.~10a). For
the total momentum $10^{24}$ g\,cm/s, the calculated maximum
amplitude corresponds to the maximum amplitude observed in the
sunquake events \cite{Kosovichev2006a}. This means that the total
momentum of the flare impact does not exceed $10^{24}$ g\,cm/s. Then
using the same solution we calculate the amplitudes of low-$l$ modes
excited by this impact. The velocity spectrum of $l=0-2$ modes is
shown in Fig.~10b. The maximum amplitude of these global modes does
not exceed 0.4 cm/s. This amplitude is about 100 times smaller than
the amplitude of stochastically excited low-$l$ modes. The maximum
amplitude of the flare-excited modes is in the frequency range of
4-5 mHz. This is also inconsistent with the suggestion of Karoff and
Kjeldsen \cite{Karoff2008} that they observed global sunquake
oscillations. Perhaps, the high-frequency frequency excess in
intensity variations is caused not by oscillations but some other
fluctuations in magnetic active regions.

In other stars the situation may be different. The energy of flares
on red dwarfs can be 3-4 orders of magnitude higher than on the Sun.
It is believed that the physics of these flares is similar to the
solar ones. Thus, if the momentum of the flare impact in the stellar
photosphere is 3 order of magnitude higher than the amplitude will
be also higher by approximately the same amount. This means that the
amplitude of global starquake waves can reach 4 m/s and thus can be
observable. The oscillations with periods of 3-5 min associated with
stellar flares have been observed (e.g. \cite{Mathioudakis2003}).
Perhaps, these were caused by the seismic response and not by
coronal loop oscillations as was suggested.

\section{Helioseismic Tomography}

Helioseismic diagnostics based on measurements and inversion of
global mode frequencies are well-known (e.g.
\cite{Kosovichev1999,JCD2002}. Most recent effort in helioseismic
applications has been focused on local techniques. Helioseismic
tomography (or time-distance helioseismology) is one of these.

\subsection{Basic Principles of Helioseismic Tomography}

The basic idea of helioseismic tomography is to measure the acoustic
travel time between different points on the solar surface, and then
to use these measurement for inferring variations of the structure
and flow velocities in the interior along the wave paths connecting
the surface points. This idea is similar to the Earth's seismology.
However, unlike in Earth, the solar waves are generated
stochastically by numerous acoustic sources in the subsurface layer
of turbulent convection. Therefore, the wave travel times are
determined from the cross-covariance function, $\Psi(\tau, \Delta)$,
of the oscillation signal, $f(t,\vecr)$, between different points on
the solar surface \cite{Duvall1993}:
\begin{equation}
\Psi(\tau,\Delta) = \int_0^T f(t,\vecr_1) f^*(t+\tau,\vecr_2)
dt,\label{EQ-104}
\end{equation}
where $\Delta$ is the angular distance between the points with
coordinates $\vecr_1$ and $\vecr_2$, $\tau$ is the delay time, and
$T$ is the total time of the observations. Because of the stochastic
nature of excitation of the oscillations, function $\Psi$ must be
averaged over some areas on the solar surface to achieve a
signal-to-noise ratio sufficient for measuring travel times $\tau$.
The oscillation signal, $f(t,\vecr)$, is usually the Doppler
velocity or intensity. A typical cross-covariance function shown in
Fig. 11  displays two sets of ridges which correspond to the first
and second  bounces of acoustic wave packets from the surface.

The cross-covariance function represents a solar `seismogram'.
Ideally, the seismogram should be inverted to infer the structure
and flows using a wave theory. However, in practice, as in
terrestrial seismology different approximations are employed, the
most simple and powerful of which is the geometrical acoustic (ray)
approximation.

Generally, the observed  solar oscillation signal corresponds to
radial displacement or pressure perturbation, and can be represented
in terms of normal modes, or standing waves:
\begin{equation}
f(t,r,\theta,\phi) = \sum_{nlm} a_{nlm}\xi_{nlm}(r,\theta,\phi)
\exp(i\omega_{nlm}t + i\phi_{nlm}),\label{EQ-105}
\end{equation}
where $n, l$ and $m$ are the radial order, angular degree and
angular order of a normal mode respectively,
$\xi_{nlm}(r,\theta,\phi)$ is a mode eigenfunction in  spherical
coordinates, $r,\theta$ and $\phi$, $\omega_{nlm}$ is the
eigenfrequency, and $\phi_{nlm}$ is an initial phase of the mode.
Using eq.~(\ref{EQ-104}), the cross-covariance function can be
expressed in terms of normal modes, and then represented as a
superposition of traveling wave packets. An example of the
theoretical cross-covariance function of p modes of the standard
solar model in Fig. 11.

By grouping the modes in narrow ranges of the angular phase
velocity, $v=\omega_{nl}/L$, where $L=\sqrt{l(l+1)}$,
 and applying the method of stationary phase,
the cross-covariance function can be approximately represented in
the form \cite{Kosovichev1997b}:
\begin{equation}
\Psi(\tau,\Delta) \propto \sum_{\delta v}
\cos\left[\omega_0\left(\tau-
\frac{\Delta}{v}\right)\right]\exp\left[-\frac{\delta\omega^2}{4}\left(\tau-
\frac{\Delta}{u}\right)^2\right],\label{EQ-106}
\end{equation}
where $\delta v$ is a narrow interval of the phase speed, $u\equiv
(\partial\omega/\partial k_h)$ is the horizontal component of the
group velocity, $k_h = 1/L$ is the angular component of the wave
vector, and $\omega_0$ is the central frequency of a Gaussian
frequency filter applied to the data, and $\delta\omega$ is the
characteristic width of this filter. Therefore, the  phase and group
travel times are measured by fitting individual terms of
eq.~({\ref{EQ-106}) to the observed  cross-covariance function using
a least-squares technique. In some cases the ridges of the
cross-covariance function may partially overlap, thus making the
interpretation of the time-distance results more difficult.

\begin{figure}
  \includegraphics[height=.5\textheight]{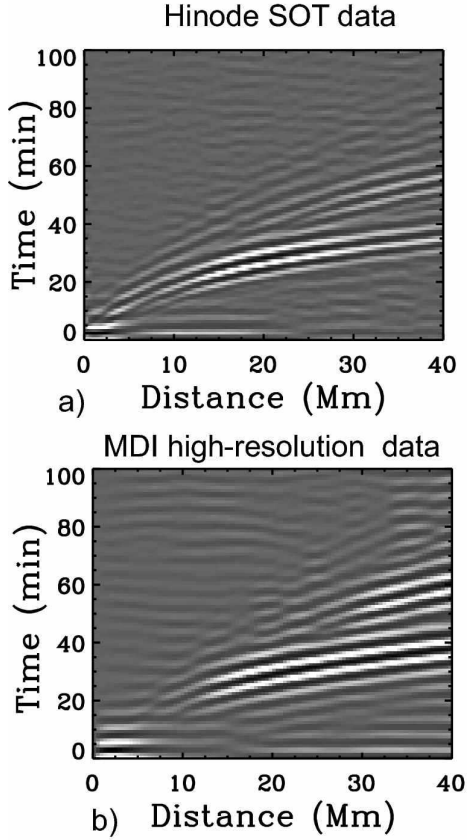}
  \caption{Cross-covariance functions of solar oscillations
  (time-distance diagrams) obtained from a) Hinode and b) MDI high-resolution
data.\citep{Sekii2007}}
\end{figure}

\begin{figure}
  \includegraphics[height=.35\textheight]{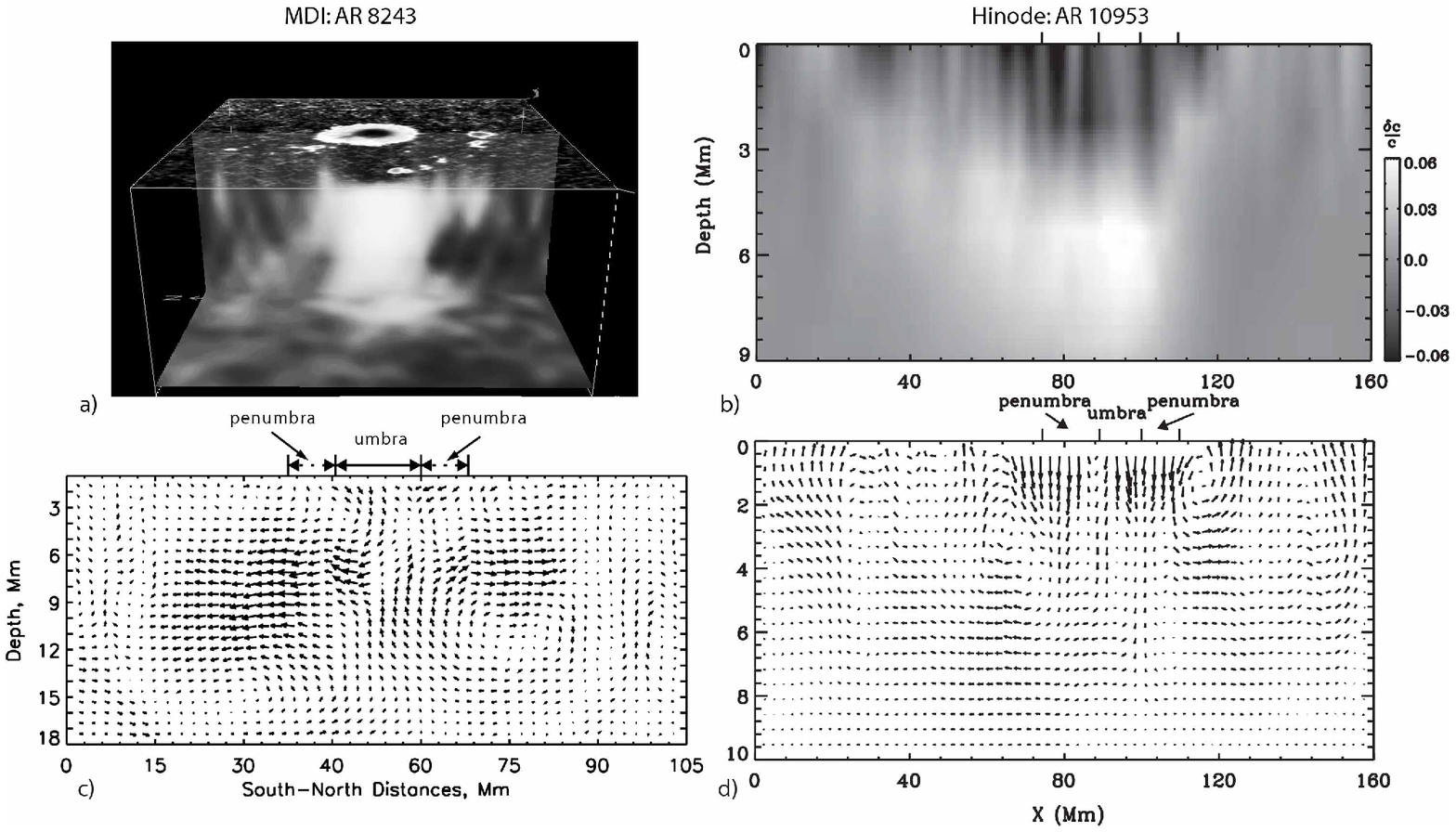}
  \caption{Wave speed perturbations($a-b$) and flow velocities
 ($c-d$) beneath sunspots from
 the MDI ($left$) and Hinode ($right$) data \citep{Zhao2009a}.}
\end{figure}

This technique measures both phase ($\Delta/v$) and group
($\Delta/u$) travel times of the p-mode wave packets.  It was found
that the noise level in the phase-time measurements is substantially
lower than in the group-time measurements. The geometrical acoustic
(ray) approximation was employed to relate the measured phase times
to the internal properties of the Sun. More precisely, the
variations of the local travel times at different points on the
surface, relative to the travel times  averaged over the observed
area are measured. Then  variations of the internal structure and
flow velocities are inferred from the travel time anomalies using a
perturbation theory.

\subsection{High-resolution Helioseismology from Hinode}

 Solar Optical Telescope (SOT) \citep{Tsuneta2008} on
Hinode spacecraft \citep{Kosugi2007}
 has unique
capabilities for high-resolution helioseismology. The
high-resolution helioseismology investigations are at the very
beginning and undoubtedly will bring new understanding of the
subsurface structure and dynamics, particularly, of the magnetized
plasma of sunspots and active regions. The unique advantage of
Hinode observation is  clearly seen by comparing the time-distance
diagrams (cross-covariance functions) obtained from Hinode and
SOHO/MDI data (Fig.~11). For the MDI data, the time-distance ridges
corresponding to signals of traveling wave packets, are not resolved
for distances shorter than 5-10 Mm. The cross-covariance signal at
these distances is dominated by horizontal artifact ridges (in the
lower left corners of Figs~11b). In the time-distance diagram
obtained from Hinode data (Fig.~11a) the wave signal is resolved for
much shorter travel distances, up to about 2 Mm. This is comparable
with the characteristic wavelength of solar oscillations.

In Figure 12 we compare subsurface structures and flows below
sunspots obtained from Hinode and MDI data. A vertical cut  along
the East-West direction approximately in the middle of a large
sunspot observed in AR 10953, May 2, 2007, (Fig.~12b), shows that
the wave speed anomalies extend about half of the sunspot size
beyond the sunspot penumbra into the plage area. In the vertical
direction, the negative wave speed perturbation extends to a depth
of 3--4 Mm. The positive perturbation is about 9 Mm deep, but it is
not clear whether it extends further, because our inversion cannot
reach deeper layers because of the small field of view. Similar
two-layer sunspot structures were observed before from SOHO/MDI
\citep{Kosovichev2000}(Fig.~12a). But, it is striking that the new
images strongly indicate on the cluster structure of the sunspot.
This was not previously seen in the tomographic images of sunspots
obtained with lower resolution. The high-resolution flow field below
the sunspot is also significantly more complicated than the
previously inferred from SOHO/MDI \citep{Zhao2001} (Fig.~12c), but
reveals the same general converging downdraft pattern. A vertical
view of an averaged flow field (Fig.~12d) shows nicely the flow
structure beneath the active region. Strong downdrafts are seen
immediately below the sunspot's surface, and extends up to 6 Mm in
depth. A little beyond the sunspot's boundary, one can find both
upward and inward flows. Clearly, large-scale mass circulations form
outside the sunspot, bringing plasma down along the sunspot's
boundary, and back to the photosphere within about twice of the
sunspot's radius. It is remarkable that such an apparent mass
circulation is obtained directly from the helioseismic inversions
without using any additional constraints, such as forced mass
conservation. Previously, the circulation pattern was not that
clear.

\section{Conclusion}

During the past decade thanks to the long-term continuous
observations from the ground and space the physics of solar
oscillations made a tremendous progress in understanding the
excitation mechanisms in various condition of the quiet Sun and
active regions, wave propagation and their interaction with magnetic
fields and turbulence, and in developing new techniques for
helioseismic diagnostics of the solar structure and dynamics.
However, many problems are still unresolved. Most of them are
related to phenomena in magnetic field regions. In particular, in
sunspots 3-min oscillations and running penumbra waves remain a
mystery. The excitation and damping mechanisms in the turbulent
magnetized plasma of active regions are far from understanding. The
processes of wave scattering and transformation must be studied in
realistic conditions of the upper convection zone and solar
atmosphere. The prime helioseismology tasks are to detect processes
of magnetic field generation and transport in the solar interior,
and formation of active regions and sunspots. This will be help to
understand the physics of the solar dynamo and the cyclic behavior
of solar activity.

For solving these tasks it is very important to continue developing
realistic MHD simulations of solar convection and oscillations and
to obtain continuous high-resolution helioseismology data for the
whole Sun. The recent observations from Hinode have convincingly
demonstrated advantages of high-resolution helioseismology, but
unfortunately such data are available only for small regions and for
short periods of time. A new substantial progress in observations of
solar oscillations is expected from the Solar Dynamics Observatory
(SDO) space mission scheduled for launch in December 2009. The
Helioseismic and Magnetic Imager (HMI) instrument on SDO will
provide uninterrupted Doppler shift measurements over the whole
visible disk of the Sun with a spatial resolution of 0.5 arcsec per
pixel ($4096\times 4096$ images) and 40-50 sec time cadence. The
total amount of data from this instrument will reach 2 Tb per day.
This tremendous amount of data will be processed through a specially
developed data analysis pipeline and will provide high-resolution
maps of subsurface flows and sound-speed structures
\citep{Kosovichev2006}. These data will enable investigations of the
multi-scale dynamics and magnetism of the Sun and also contribute to
our understanding of the Sun as a star.

%\bibliographystyle{aipproc}
%\bibliography{solar_oscillations_1}
%\end{thebibliography}
\end{document}